\documentclass[12pt]{article}
\usepackage{graphicx, amsmath, amssymb}
\usepackage{hyperref}
\usepackage{epstopdf}

\newcommand{\half}{{\frac{1}{2}}}
\newcommand{\mbf}[1]{\mathbf{#1}}

\begin{document}

\begin{flushright}
{\small
SLAC--PUB--13875\\
\date{today}}
\end{flushright}

\vspace{30pt}

\centerline{\LARGE  {Light-Front Quantization Approach to the }}

\vspace{5pt}

\centerline{\LARGE {Gauge-Gravity Correspondence and Hadron}}

\vspace{5pt}

\centerline{\LARGE  {Spectroscopy}}

\vspace{20pt}

\centerline{{
Guy F. de T\'eramond$^{a}$
%\footnote{Electronic address:gdt@asterix.crnet.cr}} 
and Stanley J. Brodsky,$^{b}$
%\footnote{Electronic address:sjbth@slac.stanford.edu}
 }}

\vspace{20pt}

{\centerline {$^{a}${Universidad de Costa Rica, San Jos\'e, Costa Rica}}

\vspace{4pt}

{\centerline {$^{b}${SLAC National Accelerator Laboratory, 
Stanford University, Stanford, CA 94309, USA}}

 \vspace{40pt}

\begin{abstract}

We find a correspondence  between semiclassical  QCD quantized on the light-front and a dual gravity model in anti--de Sitter (AdS) space, thus providing an initial approximation to QCD in its strongly coupled regime.  This correspondence --  light-front holography --  leads to a light-front Hamiltonian and relativistic  bound-state wave equations  that are functions of an invariant impact variable $\zeta$ which measures the separation of the quark and gluonic constituents within hadrons at equal light-front time.  The eigenvalues of the resulting light-front Schr\"odinger and Dirac equations are consistent with the observed light meson and baryon spectrum, and the eigenmodes provide the light-front wavefunctions, the probability amplitudes describing the dynamics of the hadronic constituents.   The  light-front equations of motion, which are dual to an effective classical gravity theory, possess remarkable algebraic and integrability properties which are dictated by the underlying conformal properties of the theory.  We  extend the algebraic construction  to include a confining potential while preserving  the integrability of the mesonic and baryonic bound-state equations.

\end{abstract}

\newpage

\section{Introduction}
Quantum Chromodynamics (QCD) provides a  fundamental description of hadrons in terms of structureless quarks and gluons.  Despite the successful applications of 
QCD in the perturbative regime, it still remains one of
the most difficult field-theoretic problems in strong interaction dynamics to understand 
the interactions and composition of hadrons in terms of the confined constituent
degrees of freedom of the QCD Lagrangian. 
Thus an important theoretical goal
is  to find an initial approximation to QCD which is 
analytically tractable and which can be systematically improved. For
example, in quantum electrodynamics, the Coulombic Schr\"odinger and
Dirac equations provide accurate first approximations to
atomic bound state problems which can then be systematically
improved using the Bethe-Salpeter formalism and correcting for
quantum fluctuations, such as the Lamb Shift and vacuum
polarization.

The AdS/CFT correspondence~\cite{Maldacena:1997re} between
string states on anti--de Sitter (AdS) space-time and conformal gauge field theories (CFT) in physical space-time has
brought a  new set of tools for studying the dynamics of strongly coupled quantum field theories, and it has led
to new analytical insights into the confining dynamics of QCD which is difficult to realize using other methods. 
The AdS/CFT duality provides a gravity
description in a ($d+1$)-dimensional AdS
space-time in terms of a flat
$d$-dimensional conformally-invariant quantum field theory defined at the AdS 
asymptotic boundary.~\cite{Gubser:1998bc}
Thus, in principle, one can compute physical observables in a strongly coupled gauge theory  in terms of a classical gravity theory. 

As first shown by Polchinski
and Strassler,~\cite{Polchinski:2001tt} the AdS/CFT duality, modified
to incorporate a mass scale, 
provides a derivation of dimensional counting
rules~\cite{Brodsky:1973kr} for the leading 
power-law fall-off of hard scattering beyond the perturbative regime.
The modified theory generates the hard behavior expected from QCD, instead of the soft
behavior characteristic of strings. In order to describe a confining theory,
 the conformal invariance of AdS$_5$ must be broken. A simple way to impose confinement and  discrete
normalizable modes is to truncate the regime where the string modes can propagate by introducing an infrared (IR) cutoff at a finite value   $z_0 \sim 1/\Lambda_{\rm QCD}$. Thus the ``hard-wall'' at $z_0$ breaks conformal invariance and allows the introduction of the QCD scale  and a spectrum of particle states.~\cite{Polchinski:2001tt}
The mechanisms of confinement are 
encoded in the warp of the AdS$_5$ metrics near a large infrared value $z_0$ which sets the scale of the strong
interactions.  In this simplified  approach
the propagation
of hadronic modes in a fixed effective gravitational background encodes salient properties of the QCD dual theory, such
as the ultraviolet (UV) conformal limit at the AdS boundary at $z \to 0$, as well as modifications of the background geometry in the 
large $z$ infrared region as are characteristic of strings dual to confining gauge theories.  
Since the AdS metric 
\begin{equation} \label{eq:AdSz}
ds^2 = \frac{R^2}{z^2} \left(\eta_{\mu \nu} dx^\mu dx^\nu - dz^2\right),
\end{equation}
is invariant under a dilatation of all coordinates $x^\mu \to \lambda x^\mu$, $z \to \lambda z$, the variable $z$ acts like a scaling variable in Minkowski space: different values of $z$ correspond to different energy scales at which the hadron is examined.

In the usual AdS/QCD approach~\cite{Erlich:2005qh, DaRold:2005zs} bulk fields are introduced to match the 
$SU(2)_L \times SU(2)_R$ chiral symmetries of QCD and its spontaneous breaking, but without explicit connection with the internal constituent structure of hadrons.~\cite{Brodsky:2003px} Instead, axial and vector currents become the
primary entities as in effective chiral theory.  Following this bottom-up approach
only a limited number of operators are introduced, and consequently only a limited
number of fields are required to construct  phenomenologically viable five-dimensional gravity duals.
One can also effectively modify the metric of AdS space by introducing a dilaton field to reproduce the observed linear Regge behavior of the hadronic spectrum.~\cite{Karch:2006pv} The resulting model is equivalent to the introduction of a harmonic oscillator confining potential $U \sim  \kappa^4 z^2$. In this  ``soft-wall'' model, the value of $\kappa$ breaks conformal invariance, and it sets the mass scale for the hadronic spectrum.  The soft-wall  background approaches asymptotically AdS space for small values of the variable $z$, as required by the near-conformal  limit of QCD in the ultraviolet region.

We have recently shown a remarkable
connection between the description of hadronic modes in AdS space and
the Hamiltonian formulation of QCD in physical space-time quantized
on the light-front (LF) at equal light-front time  $\tau$.   Light-front quantization is the ideal framework for describing the
structure of hadrons in terms of their quark and gluon degrees of
freedom. The simple structure of the light-front vacuum allows an unambiguous
definition of the partonic content of a hadron in QCD and of hadronic light-front wavefunctions
which relate its quark
and gluon degrees of freedom to their asymptotic hadronic state. The light-front wavefunctions (LFWFs) of relativistic bound states in QCD provide a fundamental description of the structure and internal dynamics of hadronic states in terms of their constituent quark and gluons  at a fixed light-front time  $\tau = t + z/c$, the time marked by the
front of a light wave,~\cite{Dirac:1949cp} --  rather than at instant time $t$, the ordinary time. Unlike instant time quantization, the Hamiltonian equation of motion in the LF is frame independent. This makes a direct connection of QCD with AdS/CFT methods possible.   

An important  first step for establishing the correspondence with AdS space is to observe that the LF bound state Hamiltonian equation of motion in QCD has an essential dependence  in the invariant transverse variable $\zeta$,~\cite{deTeramond:2008ht} which measures the
separation of the quark and gluonic constituents within the hadron
at the same LF time.  The  variable $\zeta$ plays the role of the radial coordinate $r$ in atomic systems. The result is a single-variable light-front relativistic
Schr\"odinger equation. This first approximation to relativistic QCD bound-state systems is 
equivalent to the equations of motion that describe the propagation of spin-$J$ modes in a fixed  gravitational background asymptotic to AdS space.~\cite{deTeramond:2008ht}  The eigenvalues of the LF Schr\"odinger equation give the hadronic spectrum and its eigenmodes represent the probability amplitudes of the hadronic constituents.  By using the correspondence between $\zeta$ in the LF theory and $z$ in AdS space, one can identify the terms in the dual gravity AdS equations that correspond to the kinetic energy terms of  the partons inside a hadron and the interaction terms that build confinement.~\cite{deTeramond:2008ht}  The identification of orbital angular momentum of the constituents in the light-front description is also a key element in our description of the internal structure of hadrons using holographic principles.
This mapping was originally obtained
by matching the expression for electromagnetic current matrix
elements in AdS space with the corresponding expression for the
current matrix element using LF theory in physical space
time.~\cite{Brodsky:2006uqa}  More recently we have shown that one
obtains the identical holographic mapping using the matrix elements
of the energy-momentum tensor,~\cite{Brodsky:2008pf} thus providing
a consistency test and verification of holographic
mapping from AdS to physical observables defined on the light front.

The integrability of the equations underlying a physical system is related to its
symmetries. As we shall show,  conformal symmetry in fact yields 
integrability: the differential equations describing 
the relativistic bound state system can be expressed in terms of ladder operators, and the solutions can be
written in terms of analytic functions.
The application of ladder operators~\cite{Infeld:1941} is particularly useful in the study
of AdS/QCD models~\cite{Brodsky:2008pg} and the description of
arbitrary spin fields in AdS space.~\cite{Metsaev:2008fs}
We will show how the  algebraic conformal properties can be extended
to include the mass scale of the soft-wall model and thus confinement. 
This  algebraic construction is also extended to build models of baryons.

\section{Light-Front Quantization of QCD\label{LFquant}}

One can express the  hadron four-momentum  generator $P =  (P^+, P^-, \mbf{P}_{\!\perp})$, 
$P^\pm = P^0 \pm P^3$,  in terms of the
dynamical fields, the Dirac field $\psi_+$, where $\psi_\pm = \Lambda_\pm
\psi$, $\Lambda_\pm = \gamma^0 \gamma^\pm$, and the transverse field
$\mbf{A}_\perp$ in the $A^+ = 0$ gauge~\cite{Brodsky:1997de}
quantized on the light-front at fixed light-cone time $x^+ $, $x^\pm = x^0 \pm x^3$
\begin{eqnarray} \label{eq:Pm}
P^-  &\!\!=\!&  \half \int \! dx^- d^2 \mbf{x}_\perp \bar \psi_+ \, \gamma^+   
\frac{ \left( i \mbf{\nabla}_{\! \perp} \right)^2 + m^2 }{ i \partial^+}  \psi_+
 + {\rm(interactions)} ,\\ \label{eq:Pp}
P^+ &\!\!=\!& \int \! dx^- d^2 \mbf{x}_\perp 
 \bar \psi_+ \gamma^+   i \partial^+ \psi_+ ,  \\ \label{eq:Pperp}
 \mbf{P}_{\! \perp}  &\!\!=\!&  \half \int \! dx^- d^2 \mbf{x}_\perp 
 \bar \psi_+ \gamma^+   i \mbf{\nabla}_{\! \perp} \psi_+   ,
\end{eqnarray}
where the integrals are over the initial surface $x^+ = 0$.
The LF Hamiltonian $P^-$ generates LF time translations
$\left[\psi_+(x), P^-\right] = i \frac{\partial}{\partial x^+} \psi_+(x)$,
whereas the LF longitudinal  $P^+$ and  transverse momentum $\mbf{P}_\perp$ are kinematical generators.
As in non-relativistic quantum mechanics, the LF Hamiltonian is additive in the kinetic and potential
energy. For simplicity we have omitted from (\ref{eq:Pm} - \ref{eq:Pperp}) the contributions from the gluon field $\mbf{A}_\perp$.

A physical hadron in four-dimensional Minkowski space has four-momentum $P_\mu$ and invariant
hadronic mass states $P_\mu P^\mu = M^2$ determined by the 
Lorentz-invariant Hamiltonian equation for the relativistic bound-state system
\begin{equation} \label{eq:LFH}
P_\mu P^\mu \vert  \psi(P) \rangle = \left( P^- P^+ -  \mbf{P}_\perp^2\right) \vert \psi(P) \rangle =
{M}^2 \vert  \psi(P) \rangle,
\end{equation}
where  the hadronic state $\vert\psi\rangle$ is an expansion in multiparticle Fock eigenstates
$\vert n \rangle$ of the free light-front  Hamiltonian: 
$\vert \psi \rangle = \sum_n \psi_n \vert \psi \rangle$. 
The  state $\vert \psi(P^+,\mbf{P}_\perp,J^z) \bigr\rangle$
is an eigenstate of the total momentum $P^+$
and $\mbf{P}_{\! \perp}$ and the  total spin  $J^z$. Quark and gluons appear from the light-front quantization
of the excitations of the dynamical fields $\psi_+$ and $\mbf{A}_\perp$ expanded in terms of creation and
annihilation operators at fixed LF time $\tau$. The Fock components $\psi_n(x_i, {\mathbf{k}_{\perp i}}, \lambda_i^z)$
are independent of  $P^+$ and $\mbf{P}_{\! \perp}$
and depend only on relative partonic coordinates:
the momentum fraction
 $x_i = k^+_i/P^+$, the transverse momentum  ${\mathbf{k}_{\perp i}}$ and spin
 component $\lambda_i^z$. Momentum conservation requires
 $\sum_{i=1}^n x_i = 1$ and
 $\sum_{i=1}^n \mathbf{k}_{\perp i}=0$.
The LFWFs $\psi_n$ provide a
{\it frame-independent } representation of a hadron which relates its quark
and gluon degrees of freedom to their asymptotic hadronic state.

\subsection{A Semiclassical Approximation to QCD}

We can compute $M^2$ from the hadronic matrix element
\begin{equation}
\langle \psi(P') \vert P_\mu P^\mu \vert\psi(P) \rangle  = 
M^2  \langle \psi(P' ) \vert\psi(P) \rangle,
\end{equation}
expanding the initial and final hadronic states in terms of its Fock components. The computation is  simplified in the 
frame $P = \big(P^+, M^2/P^+, \vec{0}_\perp \big)$ where $P^2 =  P^+ P^-$.
We find
 \begin{equation} \label{eq:Mk}
 M^2  =  \sum_n  \! \int \! \big[d x_i\big]  \! \left[d^2 \mbf{k}_{\perp i}\right]   
 \sum_q \Big(\frac{ \mbf{k}_{\perp q}^2 + m_q^2}{x_q} \Big)  
 \left\vert \psi_n (x_i, \mbf{k}_{\perp i}) \right \vert^2  + {\rm (interactions)} ,
 \end{equation}
plus similar terms for antiquarks and gluons ($m_g = 0)$. The integrals in (\ref{eq:Mk}) are over
the internal coordinates of the $n$ constituents for each Fock state
\begin{equation}
\int \big[d x_i\big] \equiv
\prod_{i=1}^n \int dx_i \,\delta \Bigl(1 - \sum_{j=1}^n x_j\Bigr) , ~~~
\int \left[d^2 \mbf{k}_{\perp i}\right] \equiv \prod_{i=1}^n \int
\frac{d^2 \mbf{k}_{\perp i}}{2 (2\pi)^3} \, 16 \pi^3 \,
\delta^{(2)} \negthinspace\Bigl(\sum_{j=1}^n\mbf{k}_{\perp j}\Bigr),
\end{equation}
with phase space normalization
$\sum_n  \int \big[d x_i\big] \left[d^2 \mbf{k}_{\perp i}\right]
\,\left\vert \psi_n(x_i, \mbf{k}_{\perp i}) \right\vert^2 = 1$.

The LFWF $\psi_n(x_i, \mathbf{k}_{\perp i})$ can be expanded in terms of  $n-1$ independent
position coordinates $\mathbf{b}_{\perp j}$,  $j = 1,2,\dots,n-1$, 
conjugate to the relative coordinates $\mbf{k}_{\perp i}$, with $\sum_{i = 1}^n \mbf{b}_{\perp i} = 0$.  
We can also express (\ref{eq:Mk})
in terms of the internal impact coordinates $\mbf{b}_{\perp j}$ with the result
\begin{equation}   
 M^2  =  \sum_n  \prod_{j=1}^{n-1} \int d x_j \, d^2 \mbf{b}_{\perp j} \,
\psi_n^*(x_j, \mbf{b}_{\perp j})  \\
 \sum_q   \left(\frac{ \mbf{- \nabla}_{ \mbf{b}_{\perp q}}^2  \! + m_q^2 }{x_q} \right) 
 \psi_n(x_j, \mbf{b}_{\perp j}) \\
  + {\rm (interactions)} . \label{eq:Mb}
 \end{equation}
The normalization is defined by
$\sum_n  \prod_{j=1}^{n-1} \int d x_j d^2 \mathbf{b}_{\perp j}
\left \vert \psi_n(x_j, \mathbf{b}_{\perp j})\right\vert^2 = 1$.
To simplify the discussion we will consider a two-parton hadronic bound state.  In the limit
of zero quark mass
$m_q \to 0$
\begin{equation}  \label{eq:Mbpion}
M^2  =  \int_0^1 \! \frac{d x}{x(1-x)} \int  \! d^2 \mbf{b}_\perp  \,
  \psi^*(x, \mbf{b}_\perp) 
  \left( - \mbf{\nabla}_{ {\mbf{b}}_{\perp}}^2\right)
  \psi(x, \mbf{b}_\perp) +   {\rm (interactions)}.
 \end{equation}

 The functional dependence  for a given Fock state is
given in terms of the invariant mass
\begin{equation}
 M_n^2  = \Big( \sum_{a=1}^n k_a^\mu\Big)^2 = \sum_a \frac{\mbf{k}_{\perp a}^2 +  m_a^2}{x_a}
 \to \frac{\mbf{k}_\perp^2}{x(1-x)} \,,
 \end{equation}
giving  the measure of  the off-energy shell of the bound state,
 $M^2 \! - M_n^2$.
 Similarly in impact space the relevant variable for a two-parton state is  $\zeta^2= x(1-x)\mbf{b}_\perp^2$.
Thus, to first approximation  LF dynamics  depend only on the boost invariant variable
$M_n$ or $\zeta$,
and hadronic properties are encoded in the hadronic mode $\phi(\zeta)$ from the relation
\begin{equation} \label{eq:psiphi}
\psi(x,\zeta, \varphi) = e^{i L \varphi} X(x) \frac{\phi(\zeta)}{\sqrt{2 \pi \zeta}} ,
\end{equation}
thus factoring out the angular dependence $\varphi$ and the longitudinal, $X(x)$, and transverse mode $\phi(\zeta)$
with normalization $ \langle\phi\vert\phi\rangle = \int \! d \zeta \,
 \vert \langle \zeta \vert \phi\rangle\vert^2 = 1$.
 
We can write the Laplacian operator in (\ref{eq:Mbpion}) in circular cylindrical coordinates $(\zeta, \varphi)$
and factor out the angular dependence of the
modes in terms of the $SO(2)$ Casimir representation $L^2$ of orbital angular momentum in the
transverse plane. Using  (\ref{eq:psiphi}) we find~\cite{deTeramond:2008ht}
\begin{equation} \label{eq:KV}  
M^2   =  \int \! d\zeta \, \phi^*(\zeta) \sqrt{\zeta}
\left( -\frac{d^2}{d\zeta^2} -\frac{1}{\zeta} \frac{d}{d\zeta}
+ \frac{L^2}{\zeta^2}\right)
\frac{\phi(\zeta)}{\sqrt{\zeta}}   \\
+ \int \! d\zeta \, \phi^*(\zeta) U(\zeta) \phi(\zeta) ,
\end{equation}
where all the complexity of the interaction terms in the QCD Lagrangian is summed in the effective potential $U(\zeta)$.
The LF eigenvalue equation $P_\mu P^\mu \vert \phi \rangle  =  M^2 \vert \phi \rangle$
is thus a light-front  wave equation for $\phi$
\begin{equation} \label{eq:SLFWE}
\left(-\frac{d^2}{d\zeta^2}
- \frac{1 - 4L^2}{4\zeta^2} + U(\zeta) \right)
\phi(\zeta) = M^2 \phi(\zeta),
\end{equation}
a relativistic single-variable LF Schr\"odinger equation.   Its eigenmodes $\phi(\zeta) = \langle \zeta \vert \phi \rangle$
determine the hadronic mass spectrum and represent the probability
amplitude to find $n$-partons at transverse impact separation $\zeta$,
the invariant separation between pointlike constituents within the hadron~\cite{Brodsky:2006uqa} at equal LF time. 
Extension of the results to arbitrary $n$ follows from the $x$-weighted definition of the
transverse impact variable of the $n-1$ spectator system~\cite{Brodsky:2006uqa}:
$\zeta = \sqrt{\frac{x}{1-x}} \left\vert \sum_{j=1}^{n-1} x_j \mbf{b}_{\perp j} \right\vert$, where $x = x_n$ is the longitudinal 
momentum fraction of the active quark. One can also
generalize the equations to allow for the kinetic energy of massive
quarks using Eqs. (\ref{eq:Mk}) or (\ref{eq:Mb}). In this case, however,
the longitudinal mode $X(x)$ does not decouple from the effective LF bound-state equations. 

\section{Mesons in Light-Front Holography}

On AdS space the physical  states are
represented by normalizable modes
$\Phi_P(x^\mu,z) = e^{-iP \cdot x} \Phi(z)$,
with plane waves along the Poncar\'e coordinates and a profile function $\Phi(z)$ 
along the holographic coordinate $z$. Each  LF hadronic state $\vert \psi(P) \rangle$ is dual to a normalizable string mode $\Phi_P(x^\mu,z)$  and the hadronic mass  $M^2$ is found by solving the eigenvalue problem for the corresponding wave equation in AdS space, which, as we show below, is equivalent to the semiclassical approximation Eq. (\ref{eq:SLFWE}).

As the simplest example we consider a truncated model where quark and gluons propagate freely in the
hadronic interior and the interaction terms effectively build the confinement. The effective potential is a hard wall:
$U(\zeta) = 0 ~\text{  if} ~ \zeta \le 1/\Lambda_{\rm QCD}$ and $U(z) =  \infty  ~\text{if}   ~
  \zeta > 1/\Lambda_{\rm QCD}$. This provides an analog of the MIT bag model where quarks are permanently confined inside a finite region of space.~\cite{Chodos:1974je} However, unlike bag models, boundary conditions are imposed on the 
boost invariant variable $\zeta$, not on the bag radius at fixed time.
The resulting model is a manifestly Lorentz invariant model
with confinement at large distances and conformal behavior at short distances.
Confinement is imposed with the boundary conditions
$ \phi \! \left(\zeta = 1/\Lambda_{\rm QCD}\right) = 0$.

To study the stability properties of the quantum-mechanical system, it is convenient to consider a continuous parameter $\nu$ instead of $L$ which has quantized values. If $\nu^2 \ge 0$ the  LF Hamiltonian $P_\mu P^\mu$ in the
$\zeta$-representation 
\begin{equation}
H_{LF}^\nu(\zeta)= -\frac{d^2}{d\zeta^2}
- \frac{1 - 4 \nu^2}{4\zeta^2}
\end{equation}
is written as a bilinear form
\begin{equation}
H_{LF}^\nu(\zeta) = \Pi^\dagger_\nu(\zeta) \Pi_\nu(\zeta), ~~~\nu^2 \ge 0,
\end{equation}
where the operator $\Pi_\nu$ and its adjoint  $\Pi^\dagger_\nu$ are defined by
\begin{equation} \label{eq:Pi}
\Pi_\nu(\zeta) = -i \left( \frac{d}{d \zeta} - \frac{\nu + \half}{\zeta} \right) , ~~~~
\Pi^\dagger_\nu(\zeta) = -i \left(\frac{d}{d \zeta} + \frac{\nu + \half}{\zeta}\right) ,
\end{equation}
with the commutation relation
\begin{equation}
\left[\Pi_\nu(\zeta),\Pi^\dagger_\nu(\zeta)\right] =  \frac{2 \nu+1}{\zeta^2}.
\end{equation}

For $\nu^2 \ge 0$ the Hamiltonian is positive definite,
$\langle \phi \left\vert H_{LC}^\nu \right\vert \phi \rangle
= \int d\zeta \, \vert \Pi_\nu \phi(z) \vert^2 \ge 0$,
and consequently $M^2 \ge 0$. 
For $\nu^2 < 0$ the Hamiltonian cannot be written as
a bilinear product and the Hamiltonian is unbounded from below. The
particle ``falls towards the center''~\cite{LL:1958}.
The critical value  corresponds to $\nu = 0$. Thus, if $\nu^2 < 0$ the expectation values of the Hamiltonian are
negative and $M^2 < 0$. The critical value $\nu = 0$ corresponds
the the lowest possible stable solution which we identify with the
ground state of the LF Hamiltonian and thus with zero orbital angular momentum $L=0$. 
Higher orbital excitations are constructed by the  
$L$-th application of the raising operator  on the ground state
 \begin{equation}
 \langle \zeta \vert \phi_L \rangle 
 = C_L \sqrt{\zeta} \, (-\zeta)^L \left(\frac{1}{\zeta} \frac{d}{d \zeta}\right)^L
  J_0(\zeta M) \\
  =  C_L \sqrt{\zeta} J_L\left(\zeta M\right).
  \end{equation}
The solutions $\phi_L(\zeta) = \langle \zeta \vert \phi_L \rangle $ are eigenfunctions of the LF
equation (\ref{eq:SLFWE}) with $U = 0$ and
quantized orbital excitations, $L = 0,  1,  2, \cdots$. We thus recover 
the Casimir form $L^2$ corresponding to the $SO(2)$ group of rotations in the transverse LF plane.
The mode spectrum follows from the boundary conditions at  $\zeta = 1/\Lambda_{\rm QCD}$,
and it is given in terms of the roots of the Bessel functions: 
$M_{L,k} = \beta_{L,k} \Lambda_{\rm QCD}$.  For large  quantum numbers 
$M \sim 2n + L$, in contrast to the usual Regge dependence $M^2 \sim n + L$.

Upon the substitution 
$\Phi(\zeta) \sim  \zeta^{3/2} \, \phi(\zeta)$,  $\zeta \to z$,  in the light-front wave equation (\ref{eq:SLFWE}) for $U=0$ we find
\begin{equation*} \label{eq:HWAdS}
\left[ z^2 \partial_z^2 -  3 z \, \partial_z + z^2 M^2  - (\mu R)^2\right]   \Phi(z)  = 0,
 \end{equation*}
the wave equation which describes the propagation of scalar modes in  AdS$_5$ space with
AdS radius $R$. The five-dimensional mass $\mu$ is related to the orbital angular momentum of the
hadronic bound state  by  $(\mu R)^2 = - 4 + L^2$. The quantum-mechanical stability condition $L^2 \ge 0$
is thus equivalent to the AdS Breitenlohner-Freedman
stability bound~\cite{Breitenlohner:1982jf}  $(\mu R)^2 \ge - 4$.
 Since in the conformal limit $U \to 0$, Eq. (\ref{eq:SLFWE}) is equivalent to an AdS
 wave equation for the scalar field $\Phi(z)$. The hard-wall light-front model discussed here is thus equivalent to the hard wall model of Ref.~\cite{Polchinski:2001tt}. 
 
 Each hadronic state of spin $J$ is dual to a normalizable mode 
 $\Phi_P(x^\mu,z) _{\mu_1 \cdots \mu_J} =  e^{-iP \cdot x} \Phi(z) _{\mu_1 \cdots \mu_J}$, with all the (totally symmetric) polarization indices chosen along the 3 + 1 Poincar\'e coordinates. One can thus construct an effective action in terms of
high spin modes with only the physical degrees of freedom  $\Phi(x^\mu,z) _{\mu_1 \cdots \mu_J}$ by shifting dimensions
$\Phi_J(z) =  \left(z/R\right)^{-J}  \Phi(z)$. The shifted field $\Phi_J$ obeys the equation of motion
\begin{equation} \label{eq:HWJAdS}
\left[ z^2 \partial_z^2 -  (3 \!- \!2 J) z \, \partial_z + z^2 M^2
\!  -  (\mu R)^2 \right] \!  \Phi_J = 0,
\end{equation}
where the fifth dimensional mass is rescaled according to $(\mu R)^2 \! \to (\mu R)^2 \! - J(4-J)$. Having written the wave equation
for a $J$-mode in AdS space, we can find the corresponding equation in physical 3 + 1 space
by performing the substitution $z  \to \zeta$  and
$\phi_J(\zeta)   \!  \sim \! \zeta^{-3/2 + J} \Phi_J(\zeta)$. We find
\begin{equation} \label{eq:ScheqS}
\left(-\frac{d^2}{d \zeta^2} - \frac{1-4 L^2}{4\zeta^2} \right) \phi_{\mu_1 \cdots \mu_J}
= M^2 \phi_{\mu_1 \cdots \mu_J},
\end{equation}
where $\mu$ is not a free parameter but scales according to
$ (\mu R)^2 = - (2-J)^2 + L^2$. 
In the hard-wall model there is a total decoupling of the total spin $J$ and the $\pi$ and $\rho$ mesons are degenerate.
The scaling dimensions are $\Delta = 2 + L$ independent of $J$ in agreement with the
twist scaling dimension of a two parton bound state in QCD.

\subsection{A Soft-Wall Light-Front Model for Mesons}

The conformal algebraic structure of the hard wall model
can be extended to include a scale $\kappa$. This procedure  breaks conformal
invariance and provides a solution for the confinement of modes, while maintaining an integrable algebraic structure. It also allows one to determine the stability conditions for the solutions. The resulting model resembles the soft wall model 
of Ref.~\cite{Karch:2006pv}.
We write the bound-state LF Hamiltonian as a bilinear product of operators plus a constant $C(\kappa^2)$ to be determined:
\begin{equation} \label{eq:HLC+C}
H_{LF}^\nu(\zeta) = \Pi^\dagger_\nu(\zeta) \Pi_\nu(\zeta)  + C, ~~~ \nu^2 \ge 0,
\end{equation}
where the LF generator  $\Pi$ and its adjoint $\Pi^\dagger$
\begin{equation} \label{eq:PIHO}
\Pi_\nu(\zeta) = -i \left(\frac{d}{d \zeta} - \frac{\nu + \half}{\zeta} 
- \kappa^2 \zeta \right), ~~~~
\Pi^\dagger_\nu(\zeta) = -i \left(\frac{d}{d \zeta} + \frac{\nu + \half}{\zeta}
+\kappa^2 \zeta \right),
\end{equation}
 obey the commutation relation
\begin{equation}
\left[\Pi_\nu(\zeta),\Pi^\dagger_\nu(\zeta)\right] =  \frac{2 \nu+1}{\zeta^2} - 2 \kappa^2.
\end{equation}

For $\nu^2 \ge 0$ and $C \ge - 4 \kappa^2$, the Hamiltonian is positive definite,
$\langle \phi \left\vert H_{LC}^\nu \right\vert \phi \rangle,
= \int d\zeta \, \vert \Pi_\nu \phi(z)  \vert^2+ C \ge 0$ and
$M^2 \ge 0$. For $\nu^2 < 0$ the Hamiltonian cannot be written as
a bilinear product and the Hamiltonian is unbounded from below.
The lowest stable solution of the extended LF Hamiltonian
corresponds to $C = - 4 \kappa^2$ and $\nu = 0$ and it is massless,
$M^2 = 0$.  We impose chiral symmetry by choosing $C = - 4 \kappa^2$ and thus identifying the ground state 
with the pion. With this choice of the constant $C$, the  LF Hamiltonian (\ref{eq:HLC+C}) is
\begin{equation} 
H_{LF}(\zeta) = -\frac{d^2}{d \zeta^2} 
-  \frac{1-4 L^2}{4\zeta^2} + \kappa^4 \zeta^2 + 2 \kappa^2 (L  - 1) ,
\end{equation}
with eigenfunctions
\begin{equation} \label{eq:phiSW}
\phi_L(\zeta) = \kappa^{1+L} \, \sqrt{\frac{2 n!}{(n\!+\!L\!)!}} \, \zeta^{1/2+L}
e^{- \kappa^2 \zeta^2/2} L^L_n(\kappa^2 \zeta^2),
\end{equation}
and eigenvalues $\mathcal{M}^2 = 4 \kappa^2 (n + L)$. This is illustrated in Fig. \ref{mesons} for the pseudoscalar meson spectra.

\begin{figure}
\begin{centering}
\includegraphics[angle=0,width=7.3cm]{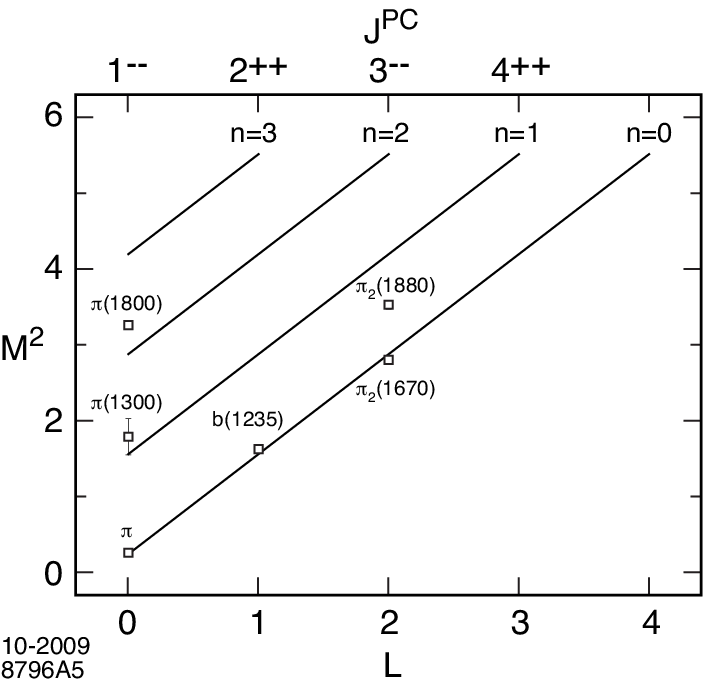} ~~~~
\includegraphics[angle=0,width=7.3cm]{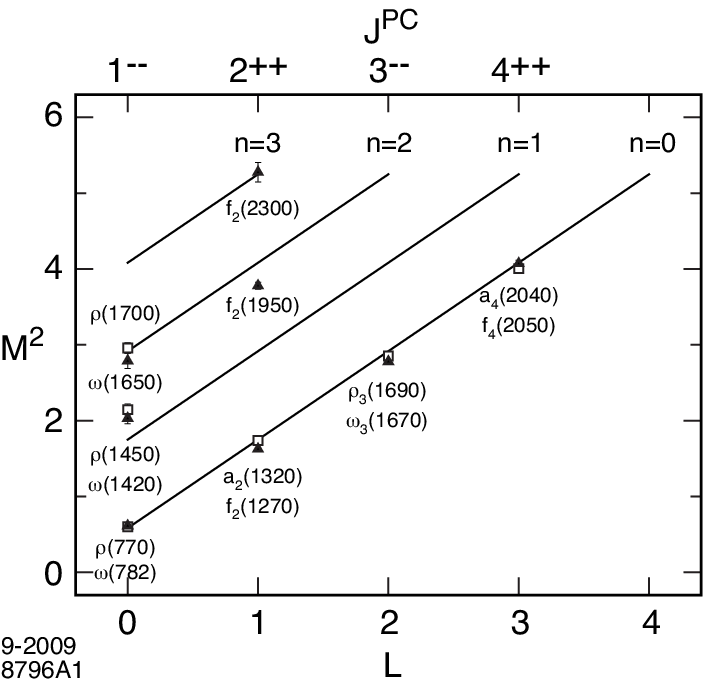}
\caption{\label{mesons} \small Parent and daughter Regge trajectories for the $\pi$-meson family for
$\kappa= 0.6$ GeV (left) and for the  $I\!=\!1$ $\rho$-meson
 and the $I\!=\!0$  $\omega$-meson families for $\kappa= 0.54$ GeV (right). The data is from \cite{Amsler:2008xx}.}
\end{centering}
\end{figure}

Next we show that the confining model has also an effective classical gravity description
corresponding to an AdS$_5$ geometry modified by a positive-sign dilaton background $exp(+ \kappa^2 z^2)$, with sign opposite
to that of reference of Ref.~\cite{Karch:2006pv}.
The positive dilaton solution has interesting physical implications, since  it leads to a confining potential between heavy quarks~\cite{Andreev:2006vy} and to a convenient framework for describing chiral
symmetry breaking.~\cite{Zuo:2009dz}  It also leads to the identification of a nonperturbative effective strong 
coupling $\alpha_s$ and 
$\beta$-functions which are in agreement with available data and  lattice 
simulations.~\cite{Deur} In presence of a dilaton profile $exp(+ \kappa^2 z^2)$ the wave equation for a spin $J$ 
mode $\Phi(z) _{\mu_1 \cdots \mu_J}$ is  given by~\cite{deTeramond:2009xx}
\begin{equation} \label{eq:SWJAdS}
\big[ z^2 \partial_z^2 - \left(3 - 2 J - 2 \kappa^2 z^2 \right) z \, \partial_z + z^2 M^2 -  (\mu R)^2 \big] \Phi_J  = 0
\end{equation}
and reduces to (\ref{eq:HWJAdS}) in the conformal limit $\kappa \to 0$.
Upon the substitution ~$z \! \to\! \zeta$  and  $\phi_J(\zeta)   \!  \sim \! \zeta^{-3/2 + J} e^{\kappa^2 \zeta^2 /2} \, \Phi_J(\zeta)$,
 we find the LF wave equation
\begin{equation}  \label{LFWEJ}
\left(-\frac{d^2}{d \zeta^2} - \frac{1-4 L^2}{4\zeta^2} 
+ \kappa^4 \zeta^2 + 2 \kappa^2(L + S - 1)  \right) \phi_{\mu_1 \cdots \mu_J} 
= M^2 \phi_{\mu_1 \cdots \mu_J},
\end{equation}
with $J_z = L_z + S_z$ and $(\mu R)^2 = - (2-J)^2 + L^2$. Equation  (\ref{LFWEJ}) has eigenfunctions
given by  (\ref{eq:phiSW}) and eigenvalues 
$M_{n, L, S}^2 = 4 \kappa^2 \left(n + L + S/2 \right)$.  The results for $S=1$ vector mesons is illustrated in Fig. \ref{mesons}, where the  spectrum is built by simply adding  $4 \kappa^2$ for a unit change in the radial quantum number, $4 \kappa^2$ for a change in one unit in the orbital quantum number and $2 \kappa^2$ for a change of one unit of spin to the ground state value of $M^2$. Remarkably, the same rule holds for baryons as shown below.

\section{Baryons in Light-Front Holography}

The effective light-front wave equation which describes baryonic states in holographic QCD is a linear equation 
determined by the LF transformation properties of spin 1/2 states.  We write
\begin{equation} \label{eq:LFD}
D_{LF}(\zeta) \psi(\zeta) = M \psi(\zeta),
\end{equation}
where  $D_{LF}$ is a hermitian operator, $D_{LF} = D^\dagger_{LF}$, thus $D_{LF}^2 = M^2$. We write $D_{LF}$ as a
product $D_{LF} = \alpha \Pi$, where $\Pi$ is the matrix valued (non-hermitian) generator
\begin{equation}
\Pi_\nu(\zeta) = -i \left( \frac{d}{d \zeta} - \frac{\nu + \half}{\zeta} \gamma \right) .
\end{equation}
If follows from the square of $D_{LF}$, $D_{LF}^2 = M^2$, that the matrices $\alpha$ and $\gamma$ are $4 \times 4$ anti-commuting hermitian matrices with unit square. The operator
$\Pi$ and its adjoint $\Pi^\dagger$ thus satisfy the commutation relation
\begin{equation}
\left[\Pi_\nu(\zeta),\Pi^\dagger_\nu(\zeta)\right] =  \frac{2 \nu+1}{\zeta^2} \, \gamma.
\end{equation}
The light front Hamiltonian $H_{LF}$ is 
\begin{equation}  
H_{LF}^\nu(\zeta) = \Pi_\nu(\zeta)^\dagger \Pi_\nu(\zeta)  
= - \frac{d^2}{d \zeta^2} 
+ \frac{\left(\nu + \half\right)^2}{\zeta^2} - \frac{\nu + \half}{\zeta^2} \, \gamma.
\end{equation}
The LF equation $H_{LF} \psi_\pm = M^2 \psi_\pm$, has a two-component solution
\begin{equation}
\psi_+(\zeta) \sim \sqrt{\zeta} J_\nu(\zeta M), ~~~~~
\psi_-(\zeta) \sim \sqrt{\zeta} J_{\nu+1}(\zeta M),
\end{equation}
where $\gamma \psi_\pm =  \pm \psi_\pm$. Thus  $\gamma$ is the four dimensional chirality operator $\gamma_5$.
In the Weyl representation
\begin{equation}
\gamma =  
  \begin{pmatrix}
  I&   0\\
  0&  -I
  \end{pmatrix}  ~~~~ {\rm and} ~~~~
 i \alpha =
  \begin{pmatrix}
  0& I\\
- I& 0
  \end{pmatrix}.
  \end{equation}
  
The effective LF equation for baryons (\ref{eq:LFD})   is indeed equivalent to the
Dirac equation describing the propagation of spin-1/2 hadronic modes,  on AdS$_5$ space
$\Psi_P(x^\mu, z) =  e^{-iP \cdot x} \Psi( z)$
\begin{equation} \label{eq:DEz}
\left[i\big( z \eta^{\ell m} \Gamma_\ell \partial_m + 2 \, \Gamma_z \big)
 + \mu R \right] \Psi = 0 ,
\end{equation}
where $\ell, m$ represent the indices of the full space with coordinates  $x^\mu$ and $z$.
Upon  the transformation $\Psi( z) \sim z^2 \psi(z)$, $z \to \zeta$,
 we recover (\ref{eq:LFD})
with $\mu R = \nu + 1/2$ and $\Gamma_z = - i \gamma$.
Higher spin fermionic modes
 $\Psi _{\mu_1 \cdots \mu_{J-1/2}}$, $J > 1/2$, with all polarization indices along the 3+1 coordinates follow by shifting
 dimensions as shown  for the case of mesons.

\subsection{A Soft-Wall Light-Front Model for Baryons}

An effective LF equation for baryons with a mass gap $\kappa$ is constructed by extending the
conformal algebraic structure for baryons described above, following the analogy with the mesons. 
We write the effective LF Dirac equation
(\ref{eq:LFD}) in terms of the matrix-valued operator $\Pi$ and its adjoint $\Pi^\dagger$ 
\begin{equation} \label{A}
\Pi_\nu(\zeta) = -i\left( \frac{d}{d \zeta} 
- \frac{\nu + \half}{\zeta} \gamma - \kappa^2 \zeta \gamma\right), ~~~
\Pi^\dagger_\nu(\zeta) =  -i\left(\frac{d}{d \zeta} 
+ \frac{\nu + \half}{\zeta} \gamma + \kappa^2 \zeta \gamma\right),
\end{equation}
with the commutation relation
\begin{equation}
\left[\Pi_\nu(\zeta),\Pi^\dagger_\nu(\zeta)\right] =  
\left(\frac{2\nu+1}{\zeta^2} - 2 \kappa^2\right) \gamma.
\end{equation}

The extended  baryonic model also has a geometric interpretation. It corresponds to the Dirac equation in AdS$_5$
space in presence of a linear potential $\kappa^2 z$
\begin{equation} \label{eq:DEz}
\left[i\big( z \eta^{\ell m} \Gamma_\ell \partial_m + 2 \, \Gamma_z \big) + \kappa^2 z
 + \mu R \right] \Psi = 0 ,
\end{equation}
as can be shown directly  by using the transformation $\Psi( z) \sim z^2 \psi(z)$, $z \to \zeta$.

Before computing the baryon spectrum we must fix the overall scale and the parameter $\nu$. As for the case of the mesons
Eq. (\ref{eq:HLC+C})
we write the LF Hamiltonian $H^\nu_{LF} = \Pi_\nu ^\dagger \Pi_\nu + C$ and chose the same value for $C$: $C = - 4 \kappa^2$.
With this choice for $C$ the LF Hamiltonian is
\begin{equation} \label{eq:LFHs}
H_{LF}= - \frac{d^2}{d \zeta^2} 
+ \frac{\left(\nu + \half\right)^2}{\zeta^2} - \frac{\nu + \half}{\zeta^2} \gamma_5 + \kappa^4\zeta^2 +
\kappa^2 (2 \nu - 3) + \kappa^2 \gamma_5.  
\end{equation}
The LF equation $H_{LF} \psi_\pm = M^2 \psi_\pm$, has a two-component solution
\begin{equation}
\psi_+(\zeta) \sim z^{\frac{1}{2} + \nu} e^{-\kappa^2 \zeta^2/2}
  L_n^\nu(\kappa^2 \zeta^2) ,\ ~~~
\psi_-(\zeta) \sim  z^{\frac{3}{2} + \nu} e^{-\kappa^2 \zeta^2/2}
 L_n^{\nu+1}(\kappa^2 \zeta^2), 
\end{equation}
and  eigenvalues $M^2 = 4 \kappa^2 (n + \nu)$, identical for both plus and minus eigenfunctions.

\begin{figure}
\begin{centering}
\includegraphics[angle=0,width=14.6cm]{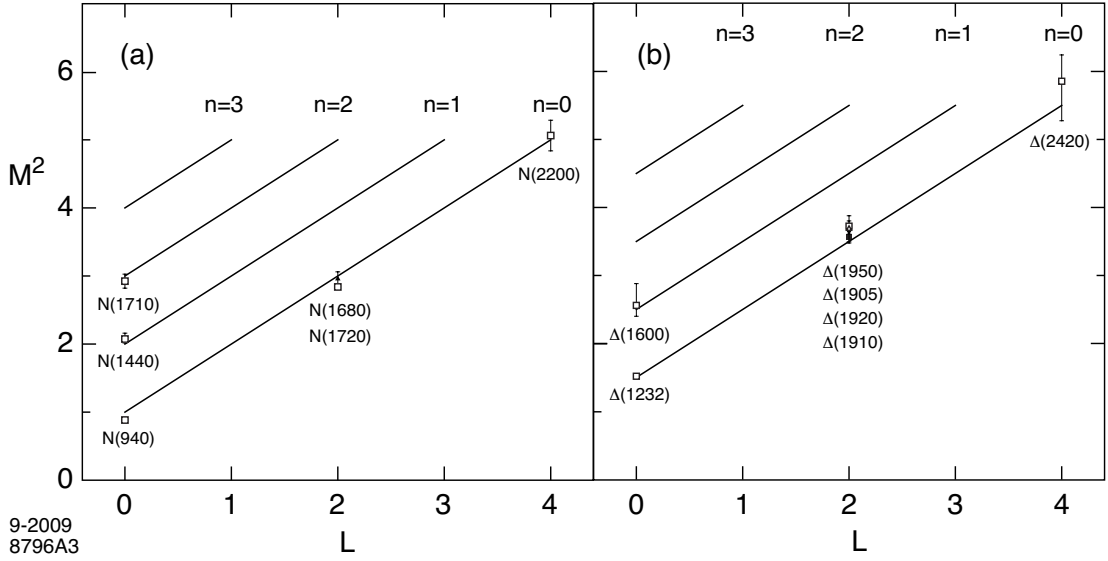}
\caption{\label{baryons}{{\bf 56} \small Parent and daughter Regge trajectories for  the  $N$ and $\Delta$ 
baryon families for $\kappa= 0.5$ GeV.
Data from \cite{Amsler:2008xx}. }
}
\label{baryons}
\end{centering}
\end{figure}

The baryon interpolating operator
$ \mathcal{O}_{3 + L} =  \psi D_{\{\ell_1} \dots
 D_{\ell_q } \psi D_{\ell_{q+1}} \dots
 D_{\ell_m\}} \psi$,  $L = \sum_{i=1}^m \ell_i$, is a twist 3,  dimension $9/2 + L$ operator with scaling behavior given by its
 twist-dimension $3 + L$. We thus require $\nu = L+1$ to match the short distance scaling behavior. Higher spin fermionic modes $\Psi _{\mu_1 \cdots \mu_{J-1/2}}$, $J > 1/2$, are obtained by shifting dimensions for the fields as in the bosonic case. 
Thus, as in the meson sector,  the increase  in the 
mass squared for  higher baryonic states is
$\Delta n = 4 \kappa^2$, $\Delta L = 4 \kappa^2$ and $\Delta S = 2 \kappa^2$, 
relative to the lowest ground state,  the proton.

The predictions for the $\bf 56$-plet of light baryons under the $SU(6)$  flavor group are shown in Fig. \ref{baryons}. As for the predictions for mesons in Fig. \ref{mesons}, only confirmed PDG~\cite{Amsler:2008xx} states are shown. 
The Roper state $N(1440)$ and the $N(1710)$ are well accounted for in this model as the first  and second radial
states. Likewise the $\Delta(1660)$ corresponds to the first radial state of the $\Delta$ family. The model is  successful in explaining the important parity degeneracy observed in the light baryon spectrum, such as the $L\! =\!2$, $N(1680)\!-\!N(1720)$ pair and the $\Delta(1905), \Delta(1910), \Delta(1920), \Delta(1950)$ states which are degenerate 
within error bars. The parity degeneracy of baryons is also a property of the hard wall model, but radial states are not well described by this model.~\cite{deTeramond:2005su} For other
recent calculations of the hadronic spectrum based on AdS/QCD, see Refs.~\cite{Boschi-Filho:2002vd,   BoschiFilho:2005yh, Evans:2006ea, Hong:2006ta, Colangelo:2007pt, Forkel:2007ru, Vega:2008af, Nawa:2008xr, dePaula:2008fp,  Colangelo:2008us, Forkel:2008un, Ahn:2009px, Sui:2009xe, Kirchbach:2009zz}.

\section{Conclusion}

We have derived a correspondence between a semiclassical first approximation to QCD quantized on the light-front
and hadronic modes propagating on a fixed AdS background. This provides a duality between the bosonic
and fermionic wave equations in AdS higher dimensional space  and the corresponding LF equations in
physical 3 + 1 space.  
The duality leads to  Schr\"odinger and Dirac-like equations  for 
hadronic bound states  in physical space-time when one identifies the 
AdS fifth dimension coordinate $z$ with the LF coordinate $\zeta$.
The light-front equations of motion, which are dual to an effective classical gravity theory, possess remarkable algebraic and integrability properties which follow from the underlying conformal properties of the theory.  We also extend the algebraic construction  to include a confining potential while preserving  the integrability of the mesonic and baryonic bound-state equations.

The light-front  holographic theory provides successful predictions for the light-quark meson and baryon spectra, as function of hadron spin, quark angular momentum, and radial quantum number.~\cite{note2} 
Using the dilaton framework the pion is massless, corresponding to zero mass quarks, in agreement with chiral invariance arguments.

Higher spin light-front equations can be derived by shifting dimensions in the AdS wave equations.
Unlike the top-down string theory approach,  one is not limited to hadrons of maximum spin
$J \le 2$, and one can study baryons with finite color $N_C=3.$

Both the hard and soft-wall models  predict similar multiplicity of states for mesons
and baryons as it is observed experimentally.~\cite{Klempt:2007cp}
In the hard-wall model the dependence  has the form:  $M \sim 2n + L$. 
However, in the soft-wall
model the observed  Regge behavior is found: $M^2 \sim n + L,$  which has the same slope in radial quantum number and orbital angular momentum. 

The semiclassical approximation to light-front QCD 
described in this paper does not account for particle
creation and absorption; it is thus expected to break down at short distances
where hard gluon exchange and quantum corrections become important. 
However, one can systematically
improve the semiclassical approximation by introducing nonzero quark masses and short-range Coulomb
corrections,  thus extending the predictions of the model to the dynamics and spectra of heavy and heavy-light quark systems.

\section*{Acknowledgments}

Invited plenary talk presented by GdT at Hadron 2009, The International Conference on Hadron Spectroscopy,
Florida State University, Tallahassee, November 29 - December 4, 2009.  We thank  Alexandre Deur, Josh Erlich and Hans-Guenter Dosch for helpful conversations and collaborations. This research was supported by the Department
of Energy contract DE--AC02--76SF00515.

\end{document}